\documentclass{amsart}

\newtheorem{theorem}{Theorem}[section]
\newtheorem{lemma}[theorem]{Lemma}

\theoremstyle{definition}

\newtheorem{xca}[theorem]{Exercise}

\theoremstyle{remark}

\numberwithin{equation}{section}

\begin{document}

\title{Set Matrices and The Path/Cycle Problem}


\author{Sergey Gubin}
\address{Genesys Telecommunication Laboratories, Inc.}
\curraddr{1255 Treat Blvd., Walnut Creek, CA 94596}
\email{sgubin@genesyslab.com}
\thanks{}


\date{September 17, 2007}

\dedicatory{}

\begin{abstract}
Presentation of set matrices and demonstration of their efficiency as a tool using the path/cycle problem.
\end{abstract}

\maketitle

\section*{Introduction}

Set matrices are matrices whose elements are sets. The matrices comprise abilities of data storing and processing. That makes them a promising combinatorial structure. To prove the concept, this work applies the matrices to the path/cycle problem, see \cite[and many others]{Tutte1, Tutte2, Ore, Cook, Karp, Chvatal, Jung, Garey, Fan, Yann, Bauer1, Diestel, TSP, Bauer2}. The problem may be generalized as a problem to find all paths and all cycles of all length in form of vertex pairs (start, finish). That is a NP-hard problem because any of its solutions will include a solution of the Hamiltonian path/cycle problems \cite{Karp}. This presentation uses set matrices to realize the following plan to solve the generalized problem: present the walk length dynamics with a generative grammar, but include in the grammar's production rules some path/cycle filters in order to deplete the resulting walk language to the indication of path/cycle's presence/absence, only. 
\newline\indent
The design's idea  may be traced back trough the dynamic programming, the Ramsey theory, the formal language theory, and to the icosian calculus \cite{Hamilton1,Hamilton2}. Realization of the design requires to maintain a set of visited/unvisited vertices and to use that set as a filter in production of the next generation of walks. Set matrices satisfy the requirements. Sorting/factoring of the visited/unvisited vertices into vertex pairs (start, finish) creates a set matrix analog of the adjacency matrix. And the especially designed powers of the set matrix create an analytic path/cycle filter. The path/cycle language's specification gets a realization in form of the easy-to-check properties of the elements of the adjacency set matrix's powers.
\newline\indent
The factoring of the set of visited/unvisited vertices into vertex pairs (start, finish) may be seen as a walk coloring where colors are the factor-sets. Then, the family of algorithms realizing the design can be parametrized with the following four extreme strategies: to color the walks with sets of the visited/unvisited start/finish vertices. Work \cite{Gubin} describes a walk coloring with the unvisited vertices. This work deploys walk coloring with the visited vertices. 
\newline\indent
Worst case for the algorithms is a complete graph. For a complete graph with $n$ vertices, the algorithms perform $n$ iterations and, on each of these iterations, $O(n^2)$-time processing for each of the $n^2$ vertex pairs. That totals in time $O(n^5)$ needed for the algorithms to find all paths and all cycles of all length in the form of vertex pairs (start, finish).

\section{Set Matrices}
\label{s:set}

Let $V$ be a universal set. Set matrices are matrices whose elements are sets. All set operations can be defined on the set matrices. For example, if $A=(a_{ij})$ and $B=(b_{ij})$ are set matrices of the appropriate sizes, then
\begin{description}
\item[Compliment]
\[
A^c = (a^c_{ij});
\]
\item[Join]
\[
A \cup B = (a_{ij}\cup b_{ij});
\]
\item[Intersection]
\[
A\cap B = (a_{ij}\cap b_{ij}),
\]
\item[Multiplication]
\[
AB = (\bigcup_\mu a_{i\mu} \times b_{\mu j}),
\]
\end{description}
- where ``$\times$'' is Cartesian product of sets, etc. More operations can be found in \cite{Gubin}.
\newline\indent
For the path/cycle problem, the most interesting operation is the set matrix multiplication. The operation can be redefined in different ways. In this presentation, let us use the following multiplication: for set matrices $A=(a_{ij})_{n\times m}$ and $B=(b_{ij})_{m\times k}$, product $AB$ is the $n\times k$ set matrix whose elements are
\begin{equation}
\label{e:product}
(AB)_{ij} = \left \{ \begin{array}{cl}
\bigcap_{\mu = 1}^{n} ~ a_{i\mu} \cup b_{\mu j}, & i \neq j \\
V, & i = j \\
\end{array} \right.
\end{equation}
Here and further, symbol $(X)_{ij}$ means $(i,j)$-element of matrix $X$. 
\newline\indent
Formula \ref{e:product} is the formula of the number matrix multiplication, except ``+'' is replaced with ``$\cap$'', ``$\times$'' is replaced with ``$\cup$'', and some special cases are taken care of. The special cases treatment makes multiplication \ref{e:product} a non-associative operation:
\begin{xca}
\label{x:1}
\[
[
\left ( \begin{array}{cc}
\emptyset & \emptyset \\
\emptyset & \{a\} \\
\end{array} \right )
\left ( \begin{array}{cc}
\emptyset & \emptyset \\
\emptyset & \{b\} \\
\end{array} \right )
]
\left ( \begin{array}{c}
\emptyset \\
\{c\}
\end{array} \right ) =
\left ( \begin{array}{cc}
V & \emptyset \\
\emptyset & V  \\
\end{array} \right )
\left ( \begin{array}{c}
\emptyset \\
\{c\} \\
\end{array} \right ) =
\left ( \begin{array}{c}
V \\
\emptyset  \\
\end{array} \right ),
\]
\[
\left ( \begin{array}{cc}
\emptyset & \emptyset \\
\emptyset & \{a\} \\
\end{array} \right )
[
\left ( \begin{array}{cc}
\emptyset & \emptyset \\
\emptyset & \{b\} \\
\end{array} \right )
\left ( \begin{array}{c}
\emptyset \\
\{c\}
\end{array} \right )] =
\left ( \begin{array}{cc}
\emptyset & \emptyset \\
\emptyset & \{a\} \\
\end{array} \right )
\left ( \begin{array}{c}
V \\
\emptyset \\
\end{array} \right ) =
\left ( \begin{array}{c}
V \\
\{a\} \\
\end{array} \right ).
\]
\end{xca}
Let $A$ be a square set matrix. The following iterations define the left and right $k$-th powers of the matrix, $k \geq 1$:
\begin{equation}
\label{e:power}
\begin{array}{cc}
R^1 = T^1 = A \\
\\
(R^{k+1})_{ij} = \left \{ \begin{array}{cl}
\bigcap_\mu (R^1)_{i\mu} \cup (R^k)_{\mu j}, & i \neq j \\
V, & i = j \\
\end{array} \right. \\
\\
(T^{k+1})_{ij} = \left \{ \begin{array}{cl}
\bigcap_\mu (T^k)_{i\mu} \cup (T^1)_{\mu j}, & i \neq j \\
V, & i = j \\
\end{array} \right. \\
\end{array}
\end{equation}
Let us estimate the computational complexity of formula \ref{e:power}. Multiplication \ref{e:product} requires $O(n^3)$ operations ``$\cup$'' and ``$\cap$''. Thus, if $t_{k-1}$ is the number of operations needed to calculate $(k-1)$-th power, then the number of operations needed to calculate $k$-th power is 
\[
t_{k} = t_{k-1} + O(n^3) = O(kn^{3}).
\]
Thus, the time needed to calculate $k$-th power can be estimated as
\begin{equation}
\label{e:estimation}
O(kn^3|V|).
\end{equation}
The list of set matrix operations and properties can be continued. But let us start and demonstrate some benefits.

\section{Path problem}
\label{s:path}

Let $g = (V,A)$ be a given (multi) digraph: $V$ is the vertex set and $A$ is the arc set of $g$. Let the vertex set $V$ be the universal set. Let's enumerate it:
\[
V = \{v_{1},~v_{2},~\ldots,~v_{n}\}.
\]
Let $G$ be the adjacency matrix of $g$ appropriate to this enumeration. Then the positive elements of powers of $G$ indicate the presence of walks: vertex pairs (start, finish) of $k$-walks are indexes of positive elements of matrix $G^k$. The powers of this adjacency matrix can detect a shortest path but not a path of a specific length. Also, calculating the powers involves magnitudes of
\[
O(n^{k-1}(\max_{ij} (G)_{ij})^k).
\]
Although, the last problem can be solved with the Boolean adjacency matrices \cite{Gubin}.
\newline\indent
Let $T$ be the following set matrix of size $n \times n$:
\begin{equation}
\label{e:T}
(T)_{ij} = \left \{ \begin{array}{cl}
\{v_{j}\}, & (G)_{ij} > 0 ~\wedge ~ i \neq j \\
V, & (G)_{ij} \leq 0 ~\vee ~ i = j \\
\end{array} \right.
\end{equation}
Matrix $T$ may be seen as an adjacency set matrix. Let $T^{k}$ be the $k$-th right power of matrix $T$, defined with formulas \ref{e:power}.
\begin{lemma}
\label{l:0}
In digraph $g$ for $k < n$, if set $(T^k)_{ij} \neq V$, then the set is equal to
\[
(T^k)_{ij} = \bigcap_\mu~\{v_{\mu_1}, v_{\mu_2}, \ldots, v_{\mu_{k-1}}, v_{\mu_k}\},
\]
where the intersection is taken over all ordered number samples 
\[
\mu = (\mu_1,\mu_2,\ldots,\mu_{k-1},\mu_k)
\]
which satisfy the following constrains:
\[
\left \{ \begin{array}{l}
1 \leq \mu_x \leq n, ~ x = 1,2,\ldots,k \\
(v_i, v_{\mu_1}) \in A, ~ (v_{\mu_x},v_{\mu_{x+1}}) \in A, ~ x = 1,2,\ldots,k-1 \\
\mu_x \neq i, ~ x = 1,2,\ldots,k \\
\mu_{k} = j \\
\mu_x \neq \mu_y ~ \Leftrightarrow x \neq y \\
\end{array} \right.
\]
- where set $A$ is the arc set of digraph $g$.
\end{lemma}
\begin{proof}
Due to definitions \ref{e:power} and \ref{e:T}, if
\[
(T^k)_{ij} = \bigcap_\mu~(T^1)_{i\mu_1} \cup (T^1)_{\mu_1\mu_2} \cup \ldots \cup (T^1)_{\mu_{k-2}\mu_{k-1}} \cup (T^1)_{\mu_{k-1}\mu_k} \neq V,
\]
then there are number samples $\mu = (\mu_1,\mu_2,\ldots,\mu_{k-1},\mu_k)$ which satisfy the first four constrains, and 
\begin{equation}
\label{e:decomposition}
(T^k)_{ij} = \bigcap_\mu~\{v_{\mu_1}\} \cup \{v_{\mu_2}\} \cup \ldots \cup \{v_{\mu_{k-1}}\} \cup \{v_{\mu_k}\},
\end{equation}
where the intersection is taken over all those number samples. Proving the last constrain will prove the lemma. To do so, let's use mathematical induction over $k$.
\newline\indent
For $k =1$, due to definitions \ref{e:power} and \ref{e:T}, $(T^1)_{ij} \neq V$ iff there are arcs from vertex $v_i$ into vertex $v_j$ and the arcs are not loops ($i \neq j$). Then, $(T^1)_{ij} = \{v_j\}$ and $(v_i, v_j) \in A$. Thus, the lemma holds for $k =1$.
\newline\indent
Because of an irregularity in the powers definition, the induction has to start from $k=2$. In this case, due to definitions \ref{e:power} and \ref{e:T}, if
\[
(T^k)_{ij} = \bigcap_\gamma~(T^1)_{i\gamma} \cup (T^1)_{\gamma j} \neq V,
\]
then there are such indexes $\gamma$ that 
\[
(T^k)_{ij} = \bigcap_{i \neq \gamma, ~ \gamma \neq j, ~ i \neq j, ~ (v_i,v_\gamma) \in A, ~ (v_\gamma, v_j) \in A}~\{v_\gamma, v_j\},
\]
where $A$ is the arc set of digraph $g$. Thus, the lemma holds for $k=2$.
\newline\indent
Let's assume that the lemma holds for all $k \leq m-1 < n-1$, and let $(T^m)_{ij} \neq V$. Then, due to decomposition \ref{e:decomposition},
\[
(T^m)_{ij} = \bigcap_\mu~\{v_{\mu_1}\} \cup \{v_{\mu_2}\} \cup \ldots \cup \{v_{\mu_{m-1}}\} \cup \{v_{\mu_m}\} \neq V,
\]
where the intersection is taken over some number samples $\mu$, satisfying the first four constrains. Then, there is such number sample $\mu$ that
\[
\{v_{\mu_1}\} \cup \{v_{\mu_2}\} \cup \ldots \cup \{v_{\mu_{m-1}}\} \cup \{v_{\mu_m}\} = Z \neq V.
\]
Then, due to decomposition \ref{e:decomposition}, for any of such number samples $\mu$, the following holds:
\[
(T^{m-1})_{i\mu_{m-1}} \subseteq \{v_{\mu_1}\} \cup \{v_{\mu_2}\} \cup \ldots \cup \{v_{\mu_{m-1}}\} \subseteq Z \neq V,
\]
and
\[
(T^{m-1})_{\mu_1\mu_{m}} \subseteq \{v_{\mu_2}\} \cup \{v_{\mu_3}\} \cup \ldots \cup \{v_{\mu_{m}}\} \subseteq Z \neq V.
\]
Then, due to the induction hypothesis, both number samples
\[
(\mu_1,\mu_2,\ldots,\mu_{m-2},\mu_{m-1})
\]
and
\[
(\mu_2,\mu_3,\ldots,\mu_{m-1},\mu_m)
\]
satisfy all five constrains. Particularly,
\[
\mu_x \neq \mu_y, ~ \Leftrightarrow ~ x \neq y, ~ x,y = 1,2,\ldots,m-1;
\]
\[
\mu_x \neq \mu_y, ~ \Leftrightarrow ~ x \neq y, ~ x,y = 2,3,\ldots,m;
\]
and, due to the third constrain for $(T^{m-1})_{\mu_1\mu_{m}} \neq V$,
\[
\mu_1 \neq \mu_m = j.
\]
Thus, the whole number sample $\mu$ satisfies the fifth constrain. That concludes the induction and proves the lemma for all $k < n$.
\end{proof}
Lemma \ref{l:0} allows the following interpretation:
\begin{lemma}
\label{l:1}
In digraph $g$, if $(T^k)_{ij} \neq V$, then there is a $k$-path from vertex $v_i$ into vertex $v_j$. 
\end{lemma}
\begin{proof}
The constrains in lemma \ref{l:0} are the definition of a path from $v_i$ into $v_j$.
\end{proof}
Lemmas \ref{l:0} and \ref{l:1} show that matrices $T^k$ collect the vertex-bridges. That may be interesting for the graph toughness theory \cite{Chvatal, Bauer2}.
\begin{lemma}
\label{l:2}
In digraph $g$, if there is a $k$-path from vertex $v_i$ into vertex $v_j$ then $(T^{k})_{ij} \neq V$.
\end{lemma}
\begin{proof}
Let the following vertices constitute a $k$-path from vertex $v_i$ into vertex $v_j$:
\[
v_{\mu_1 = i}, ~ v_{\mu_2}, ~ \ldots, ~ v_{\mu_{k+1} = j}.
\]
Indexes of these vertices satisfy the constrains in lemma \ref{l:0}. Then, due to definitions \ref{e:power} and \ref{e:T}, 
\[
(T^k)_{ij} = \bigcap_\mu~(T^1)_{i\mu_1} \cup (T^1)_{\mu_1\mu_2} \cup \ldots \cup (T^1)_{\mu_{k-2}\mu_{k-1}} \cup (T^1)_{\mu_{k-1}\mu_k} \subseteq 
\]
\[
\subseteq \{v_{\mu_1}\} \cup \{v_{\mu_2}\} \cup \ldots \cup \{v_{\mu_k}\} \cup \{v_{\mu_{k+1}}\} \subseteq V - \{v_i\}  \neq V.
\] 
\end{proof}
\begin{theorem}
\label{t:1}
In digraph $g$ for $k \geq 1$, there are $k$-paths from vertex $v_i$ into vertex $v_j$ iff 
\[
(T^k)_{ij} \neq V.
\]
\end{theorem}
\begin{proof}
The theorem aggregates lemmas \ref{l:1} and \ref{l:2}. Let us notice that case $k \geq n$ is covered by lemmas \ref{l:0} and \ref{l:2}:
\[
k \geq n ~ \Rightarrow ~ T^k = (V)_{n\times n}.
\]
\end{proof}
Estimation \ref{e:estimation} shows the computational complexity to detect the $k$-paths with theorem \ref{t:1}. Particularly, when $k=n-1$, the theorem detects the existence or absence of Hamiltonian paths in time
\[
O(n^5).
\]
\indent
All the results can be repeated with the left powers of matrix $T$. Also, definition \ref{e:T} uses the arc finish vertices. Obviously, the results can be repeated with the start vertices using the following set matrix instead of matrix \ref{e:T}:
\begin{equation}
\label{e:R}
(R)_{ij} = \left \{ \begin{array}{cl}
\{v_{i}\}, & (G)_{ij} > 0 ~\wedge ~ i \neq j \\
V, & (G)_{ij} \leq 0 ~\vee ~ i = j \\
\end{array} \right.
\end{equation}
Colorings \ref{e:T} and \ref{e:R} cover two of the four extreme strategies of walk coloring: to color walks with the visited start/finish vertices. Another two extreme strategies are discussed in \cite{Gubin}. They produce the same results but in terms of the compliment sets.

\section{Cycle problem}
\label{s:cycle}

Obviously, the solution of the path problem described in section \ref{s:path} solves the cycle problem, as well. Let us formalize that analytically.
\newline\indent
Let's define another set matrix multiplication: if $A$ and $B$ are set matrices of appropriate sizes, then
\begin{equation}
\label{e:product2}
(AB)_{ij} = \left \{ \begin{array}{cl}
\bigcap_\nu~(A)_{i\nu} \cup (B)_{\nu j}, & i = j \\
V, & i \neq j \\
\end{array} \right.
\end{equation}
And let us define the following walk coloring:
\[
(S)_{ij} = \left \{ \begin{array}{cl}
\{\mbox{``Loop''}\}, & (G)_{ij} > 0 ~\wedge ~ i = j \\
V, & (G)_{ij} \leq 0 ~\vee ~ i \neq j \\
\end{array}, \right.
\]
\begin{equation}
\label{e:Sk}
S^1 = S, ~ S^{k+1} = T^k R^1, ~ k \geq 1,
\end{equation}
- where set matrices $T^k$ and $R^1$ were defined in section \ref{s:path}, and matrix multiplication \ref{e:product2} is used.
\begin{theorem}
\label{t:2}
In digraph $g$ for $k \geq 1$, there are $k$-cycles attached to vertex $v_i$ iff
\[
(S^k)_{ii} \neq V.
\]
\end{theorem}
\begin{proof}
Case when $k=1$ is obvious. Let $k > 1$.
\newline\indent
Necessity. Let a $k$-cycle be attached to vertex $v_i$, and let the cycle visit the following vertices in the order shown:
\[
v_{\mu_1=i},~ v_{\mu_2}, ~\ldots,~ v_{\mu_{k}}, ~ v_{\mu_{k+1} = i}.
\]
Then, the last $k$ vertices in the row constitute a $(k-1)$-path from $v_{\mu_2}$ into $v_i$. Thus, due to lemma \ref{l:0} and theorem \ref{t:1}, 
\[
v_i \in (T^{k-1})_{\mu_2 i} \neq V.
\]
On the other hand, due to definition \ref{e:R},
\[
(R^1)_{i\mu_2} = \{v_i\} \neq V.
\]
Thus, due to definition \ref{e:product2},
\[
(S^k)_{ii} = ((T^{k-1})_{\mu_2 i} \cup  \{v_i\}) ~ \cap \ldots ~\subseteq ~(T^{k-1})_{\mu_2 i}\cup \{v_i\} = (T^{k-1})_{\mu_2 i} ~\neq ~ V.
\]
\indent
Sufficiency. Let
\[
(S^k)_{ii} = \bigcap_\nu (T^{k-1})_{i \nu} \cup (R^1)_{\nu i}  \neq V.
\]
Then, there is such number $\nu$ that
\[
(T^{k-1})_{i\nu}\cup (R^1)_{\nu i}  \neq V.
\]
Then, due to theorem \ref{t:1}, there is a $(k-1)$-path from $v_i$ into $v_\nu$; and, due to definition \ref{e:R}, there is an arc from $v_\nu$ into $v_i$. The path and arc create a $k$-cycle attached to $v_i$.
\end{proof}
Estimation \ref{e:estimation} gives the computational complexity of theorem \ref{t:2}. Particularly, when $k = n$, the theorem detects the existence/absence of Hamiltonian cycles in time $O(n^5)$. But some simplifications are possible. The existence/absence of Hamiltonian cycles can be detected by only calculating any one string of matrix $T^{n-1}$. That reduces the time needed to solve the Hamiltonian cycle problem to
\[
O(n^4).
\]

\section*{Conclusion}
The paper presented set matrices as an efficient tool for solving the combinatorial problems. The matrices were used to solve the path/cycle problem in polynomial time:
\begin{description}
\item[$k$-path]
Calculate set matrix $T^k$ with formulas \ref{e:T} and \ref{e:power}. Use theorem \ref{t:1} to detect all $k$-paths in form vertex pair (start, finish);
\item[$k$-cycle]
Calculate set matrix $S^k$ with formulas \ref{e:T}, \ref{e:power}, \ref{e:R}, \ref{e:product2}, and \ref{e:Sk}. Use theorem \ref{t:2} to detect all vertices which have a $k$-cycle attached.
\end{description}
Boolean property ``It is equal to the vertex set'' of the elements of matrices $T^k$ and $S^k$ fulfill the path/cycle language's specification: indicate the presence/absence of paths/cycles. For a graph with $n$ vertices, it will take $O(n^5)$-time to write down the whole language in form of $O(n)$ matrices of size $n\times n$ filled with $1$ and $0$: $1$ will mean the existence of appropriate paths/cycles and $0$ will mean their absence.


\end{document}